\documentclass[prc,aps,twocolumn,showpacs,nofootinbib]{revtex4}
\usepackage{graphicx}
\usepackage{nicefrac}
\usepackage{amsmath}
\usepackage{subfigure}
\def\pcomma{$^,$}
\def\pedinb{$^1$}
\def\pmainz{$^2$}
\def\pglsg{$^3$}
\def\pkent{$^4$}
\def\pbonn{$^5$}
\def\pgatch{$^6$}
\def\pgiess{$^7$}
\def\ppavia{$^8$}
\def\pgwu{$^9$}
\def\pucla{$^{10}$}
\def\plpi{$^{11}$}
\def\psackv{$^{12}$}
\def\pbasel{$^{13}$}
\def\pinr{$^{14}$}
\def\pzagreb{$^{15}$}
\def\pcua{$^{16}$}

\begin{document}

\title{Measurement of the $^{1}H$($\vec{\gamma}$, 
	$\vec{p}$)$\pi^{0}$ reaction using a novel 
	nucleon spin polarimeter}
\author{M.~H.~Sikora\pedinb\footnote[1]{Present address:
        The George Washington University, USA},  
	D.~P.~Watts\pedinb,
	D.~I.~Glazier\pedinb,     
        P.~Aguar-Bartolom\'e\pmainz,
        L.~K.~Akasoy\pmainz,
        J.~R.~M.~Annand\pglsg,
        H.~J.~Arends\pmainz,
        K.~Bantawa\pkent,
        R.~Beck\pbonn,
        V.~S.~Bekrenev\pgatch,
        H.~Bergh\"auser\pgiess,
        A.~Braghieri\ppavia,
        D.~Branford\pedinb,
	W.~J.~Briscoe\pgwu,
        J.~Brudvik\pucla,
        S.~Cherepnya\plpi,
        R.~F.~B.~Codling\pglsg,
        B.~T.~Demissie\pgwu,
        E.~J.~Downie\pmainz\pcomma\pglsg\pcomma\pgwu,
        P.~Drexler\pgiess,
        L.~V.~Fil'kov\plpi,
        B.~Freehart\pgwu,
        R.~Gregor\pgiess,
        D.~Hamilton\pglsg,
        E.~Heid\pmainz\pcomma\pgwu,
        D.~Hornidge\psackv,
        I.~Jaegle\pbasel,
        O.~Jahn\pmainz,
        T.~C.~Jude\pedinb,
        V.~L.~Kashevarov\plpi,
        I.~Keshelashvili\pbasel,
        R.~Kondratiev\pinr,
        M.~Korolija\pzagreb,
        M.~Kotulla\pgiess,
        A.~A.~Koulbardis\pgatch,
        S.~P.~Kruglov\pgatch,
        B.~Krusche\pbasel,
        V.~Lisin\pinr,
        K.~Livingston\pglsg,
        I.~J.~D.~MacGregor\pglsg,
        Y.~Maghrbi\pbasel,
        D.~M.~Manley\pkent,
        Z.~Marinides\pgwu,
        M.~Martinez\pmainz,
        J.~C.~McGeorge\pglsg,
        B.~McKinnon\pglsg,
        E.~F.~McNicoll\pglsg,
        D.~Mekterovic\pzagreb,
        V.~Metag\pgiess,
        S.~Micanovic\pzagreb,
        D.~G.~Middleton\psackv,
        A.~Mushkarenkov\ppavia,
        B.~M.~K.~Nefkens\pucla,
        A.~Nikolaev\pbonn,
        R.~Novotny\pgiess,
        M.~Ostrick\pmainz,
        P.~B.~Otte\pmainz,
        B.~Oussena\pmainz\pcomma\pgwu,
        P.~Pedroni\ppavia,
        F.~Pheron\pbasel,
        A.~Polonski\pinr,
        S.~Prakhov\pucla,
        J.~Robinson\pglsg,
        G.~Rosner\pglsg,
        T.~Rostomyan\ppavia\footnote[2]{Present address:
           University of Basel, Switzerland},
        S.~Schumann\pmainz,
        D.~I.~Sober\pcua,
        A.~Starostin\pucla,
	I.~I.~Strakovsky\pgwu,
        I.~M.~Suarez\pucla,
        I.~Supek\pzagreb,
        M.~Thiel\pgiess,
        A.~Thomas\pmainz,
        M.~Unverzagt\pmainz,
        D.~Werthm\"uller\pbasel,
	R.~L.~Workman\pgwu,
        I.~Zamboni\pzagreb, and
        F.~Zehr\pbasel\\
\vspace*{0.1in}
(A2 Collaboration at MAMI)
\vspace*{0.1in}
}
\affiliation{
\pedinb SUPA, School of Physics, University of Edinburgh, Edinburgh EH9 3JZ, UK}

\affiliation{
\pmainz Institut f\"ur Kernphysik, University of Mainz,
        D-55099 Mainz, Germany}

\affiliation{
\pglsg SUPA, School of Physics and Astronomy, University
        of Glasgow, Glasgow G12 8QQ, UK}

\affiliation{
\pkent Kent State University, Kent, Ohio 44242, USA}

\affiliation{
\pbonn Helmholtz-Institut f\"ur Strahlen- und Kernphysik,
        University of Bonn, D-53115 Bonn, Germany}

\affiliation{
\pgatch Petersburg Nuclear Physics Institute, 188300
        Gatchina, Russia}

\affiliation{
\pgiess II Physikalisches Institut, University of Giessen,
        D-35392 Giessen, Germany}

\affiliation{
\ppavia INFN Sezione di Pavia, I-27100 Pavia, Italy}

\affiliation{
\pgwu  The George Washington University, Washington, DC 20052, USA}

\affiliation{
\pucla University of California Los Angeles, Los Angeles,
    California 90095-1547, USA}

\affiliation{
\plpi Lebedev Physical Institute, 119991 Moscow, Russia}

\affiliation{
\psackv Mount Allison University, Sackville, New Brunswick
    E4L3B5, Canada}

\affiliation{
\pbasel Department Physik, University of Basel,
        CH-4056 Basel, Switzerland}

\affiliation{
\pinr Institute for Nuclear Research, 125047 Moscow,
        Russia}

\affiliation{
\pzagreb Rudjer Boskovic Institute, HR-10000 Zagreb,
        Croatia}

\affiliation{
\pcua The Catholic University of America, Washington,
    DC 20064, USA\\}

\date{\today}

\begin{abstract}
We report the first large-acceptance measurement of polarization 
transfer from a polarized photon beam to a recoiling nucleon, 
pioneering a novel polarimetry technique with wide application 
to future nuclear and hadronic physics experiments. The 
commissioning measurement of polarization transfer in the 
$^{1}H$($\vec{\gamma}$,$\vec{p}$)$\pi^{0}$ reaction in the range 
$0.4<E_{\gamma}<1.4$~GeV is highly selective regarding the basic 
parameterizations used in partial wave analyses to extract the 
nucleon excitation spectrum. The new data strongly favor the recently 
proposed Chew-Mandelstam formalism.  
\end{abstract}

\maketitle

\section{Introduction}
\label{sec:intro}

Spin polarization observables are a powerful tool in nuclear and 
hadronic physics, providing essential constraints on the dynamics 
of strongly bound systems and ultimately non-perturbative QCD. 
Previous measurements of nucleon spin polarization have been 
limited by small detector acceptances, resulting in the need 
for long beam times and sequential experimental measurements. 
This is generally due to the polarimeters employed in such 
experiments, which rely on accurately measuring the nucleon 
momentum before and after a spin-dependent nucleon-nucleus 
scattering interaction using charged particle tracking detectors. 
The high cost of these detector systems restricts the solid 
angular coverage.

In this letter, we present a novel approach to nucleon polarimetery 
that achieves a determination of the spin directions of protons 
with large acceptance, a goal that has remained elusive for many 
decades. The technique utilizes a reconstruction of the kinematics 
of the nucleon-nucleus scattering processes in the analyzing medium 
without the need for tracking detectors. Detailed polarized particle tracking 
simulations built on the Geant4~\cite{geant4} framework are used to isolate and 
characterize the analyzing reaction. The polarimeter concept 
presented here has the potential to provide spin-polarization 
data for a wide range of future hadronic and nuclear physics experiments 
including single and multiple meson photoproduction, 
deuteron photodisintegration and Deeply Virtual Compton Scattering. 
The polarimeter will provide neutron spin transfer observables 
in future approved experiments~\cite{proposal}. 
The commissioning reaction for the polarimeter is taken as the 
beam-recoil double polarization observable $C_x$$^\ast$ in the 
$^{1}H(\vec{\gamma},\vec{p})\pi^{0}$ reaction. This observable 
measures the degree of circular polarization transferred from the 
incident photon beam to the recoiling nucleon. There are sparse but 
accurate data for this reaction obtained at Jefferson Lab (JLab) 
using the spectrometer facilities at Hall~A~\cite{Gilman} and 
Hall~C~\cite{Perdrisat}, which can be used as a verification 
of the new polarimeter concept. 

Measurements of beam-recoil observables in meson photoproduction 
with large acceptance are a pre-requisite for improving our 
knowlege of the excitation spectrum of the nucleon, one of the 
highest priority programs in hadronic physics. A rich spectrum 
of excited states are expected for the nucleon, reflecting its 
composite nature as a tightly bound strongly interacting system 
comprising valence quarks, sea quarks, and gluons. Recently, 
theoretical predictions of this spectrum have emerged directly 
from Quantum Chromodynamics (QCD) via the lattice~\cite{lattice}, 
Holographic Dual~\cite{holo}, and Dyson-Schwinger approaches~\cite{dyson}. 
These reveal the spectrum as a sensitive test of non-perturbative 
QCD as a complete theory to describe light quark bound systems. 
The spectrum also guides phenomenological QCD based approaches 
such as constituent quark models~\cite{CQM}. 

To extract information on the excitation spectrum, cross sections 
and polarization observables measured in meson production data 
are fit with Partial Wave Analyses (PWA), based on amplitudes 
that include couplings to intermediate nucleon resonant states 
as well as non-resonant background terms. The current determination 
of the nucleon excitation spectrum is incomplete with masses, 
lifetimes, and widths of many states not conclusively measured. 
Even the existence of some states is in doubt due to contradictory 
results between different PWA~\cite{PWArev}. Accurately 
constraining PWA requires at least seven appropriately chosen cross 
section and polarization observables~\cite{CGLN} to determine 
the reaction amplitudes up to an overall phase. This has led to 
a significant effort to produce polarized nucleon targets at 
ELSA~\cite{ELSA}, MAMI~\cite{Mainz} and JLab~\cite{FROST}. However, a full constraint on PWA 
necessitates large-acceptance measurements of double polarization observables 
including recoil polarization~\cite{Ron,Chiang}. 

\begin{figure}
\includegraphics[width=1\columnwidth]{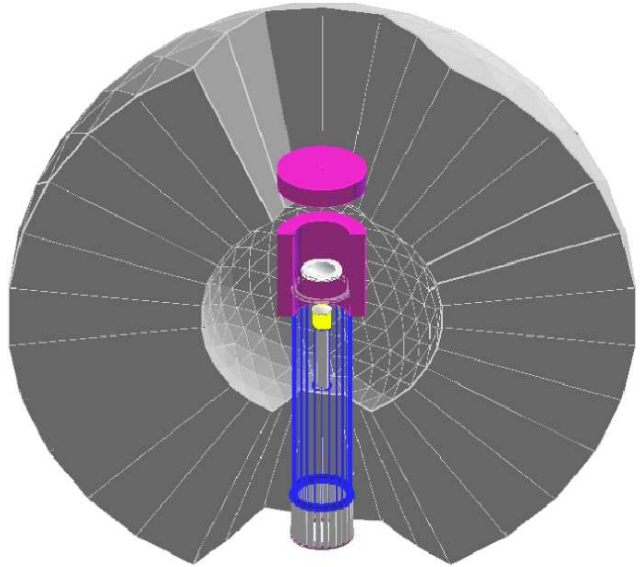}
\caption{(Color online) Illustration of the experimental setup. 
	The liquid hydrogen target (yellow cylinder), situated at 
	the center of the CB, is surrounded by the PID (blue) and 
	the 2.25~cm thick graphite polarimeter. The 7.25~cm thick 
	upstream cap covers the upstream aperture to TAPS (not 
	pictured). The PID was flush with the upstream cap during 
	data acquisition, however for this picture, the PID has 
	been shifted for clarity. \label{fig:polpic}}
\end{figure}

\section{Experimental Details}
\label{sec:DB}

This experiment took place at the Mainz Microtron (MAMI) electron 
accelerator facility~\cite{Arends,Walcher} in a total beamtime of 
600 hours. Circularly polarized bremsstrahlung photons 
were energy tagged by the Glasgow Tagger~\cite{HallTagg,McGTagg} and 
impinged on a 5~cm long liquid hydrogen target. Reaction products were 
detected with the Crystal Ball (CB)~\cite{CBOreglia,CBHist,CBStarostin}, 
a highly segmented NaI(Tl) photon calorimeter covering nearly 96\% of 
4$\pi$, and the TAPS~\cite{TAPSNovotny,Gabler} BaF$_2$ array, which covered 
the forward angle region for $\theta = 5-20^{\circ}$. The 
experimental apparatus is shown schematically in Fig.~\ref{fig:polpic}. 
A 24 element 50~cm long scintillator barrel (PID) surrounded the 
target to assist in charged particle identification. For this 
experiment, additional analyzing material for the polarimeter was 
placed inside the CB, comprising a 2.5~cm thick graphite cylinder 
covering $\theta\ge 12^\circ$ placed outside the PID and a 7.25~cm 
thick upstream cap covering $\theta < 12^\circ$.   

The events of interest are those for which the proton has undergone 
a nuclear scatter with a $^{12}$C nucleus in the polarimeter. The 
successful operation of the polarimeter requires such events to be 
cleanly identified for a wide range of incident proton angles and 
energies. The analysis utililizes a kinematic reconstruction of the 
scattering of final state protons in the graphite analyzer, exploiting 
the accurate measurement of the incident photon, final state pion and 
the interaction point of the proton in the CB. For the 
$^{1}H(\vec{\gamma},\vec{p})\pi^{0}$ events, three particles 
are detected in the final state: the proton and the two photons from 
the fast $\pi^{0}\rightarrow 2\gamma$ decay. 
The two particles whose invariant mass was closest to the $\pi^0$ mass of 
135~MeV were identified as photons, with an additional check for anti-coincidences in the PID. A missing mass cut was then used to identify the recoiling 
proton and to reconstruct its 4-vector ${\bf p_{rec}}$. A correlation 
in the azimuthal angle of the reconstructed proton and a hit in the 
appropriate PID element were also required. Particle identification 
was confirmed using $\Delta E$-$E$ analysis of the reconstructed 
proton energy with the PID energy signal offering clean identification 
up to proton energies of 650~MeV, well above the energy at which protons 
are stopped within the CB crystals. 

The spin-orbit term in the nucleon-nucleus potential introduces an 
azimuthal modulation for the scattered proton given by~\cite{Wolf,Wolf2}
\begin{equation}
\begin{split}
	&N(\theta_{sc},\phi_{sc}) = \\
        &N_0(\theta_{sc}) \left [1 
	+ A(\theta_{sc})(P 
	\mbox{cos}\phi_{sc} - C_x^{\ast}P_{\gamma}^{\odot} 
	\mbox{sin}\phi_{sc}) \right ]
\label{eq:scat}
\end{split}
\end{equation}
$\theta_{sc}$ and $\phi_{sc}$ describe the scattering of 
the protons off the carbon nuclei in a lab 
coordinate frame defined by ${\bf z'}$ along the initial proton momentum 
${\bf p_{rec}}$; ${\bf y'}$ along $\bf{k_{\gamma} \times k_{\pi}}$; 
and ${\bf x'}$ along $\bf{y' \times z'}$.
$N_0(\theta_{sc})$ is the unpolarized scattering distribution, 
$A(\theta_{sc})$ is the p-$^{12}$C analyzing power, $P$ is the single 
polarization observable describing the induced polarization in the 
y-direction of the scattering frame, and $P_{\gamma}^{\odot}$ is the 
degree of circular polarization of the photon beam. 
To determine the scattering angles, the hit position in the graphite 
and CB were required. The latter was determined by averaging the positions
of the struck CB crystals, weighted by the energy in each cystal. The former was approximated by assuming 
the proton with momentum ${\bf p_{rec}}$ initiated from the center 
of the target and scattered at the mid-point of the graphite scatterer.
The vector between the two positions was then rotated into the 
coordinate frame described above. 
\begin{figure}
\subfigure{
\includegraphics[width=1\columnwidth]{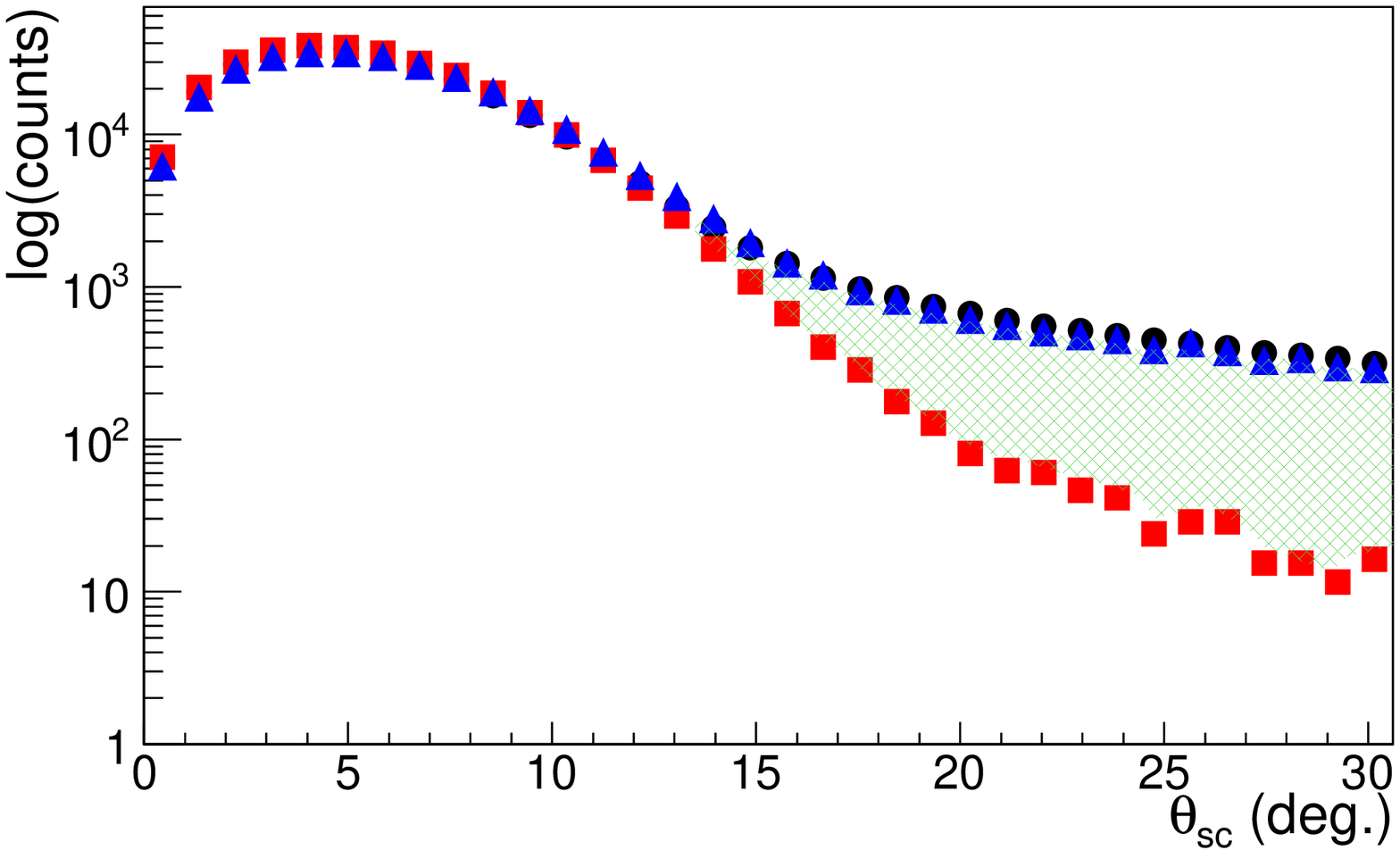}
}
\subfigure{
\includegraphics[width=1\columnwidth]{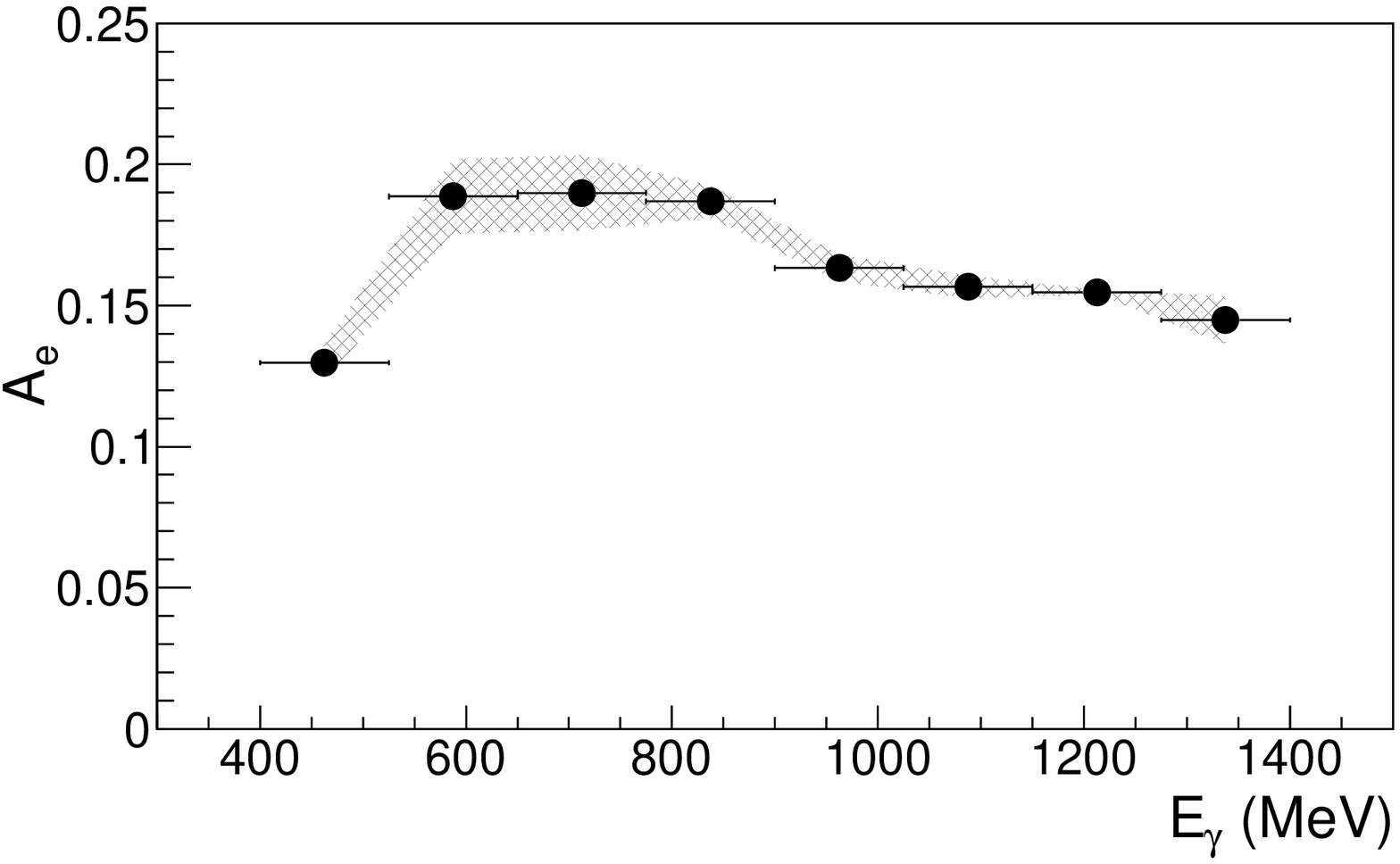}
}
\caption{(Color online) (Top) Comparison of $\theta_{sc}$ for data (circles), 
	simulation (triangles), and the simulation with no hadronic 
	interaction (squares). Nuclear scattered events lie in the 
	shaded region. (Bottom) The average of the effective analyzing powers
        obtained from each model, integrated over $\theta_{CM}$. The shaded region indicates the systematic error.
        \label{fig:comp}}
\end{figure}

During the experiment, the helicity of the photon was flipped 
randomly every second, and an asymmetry was formed between the 
azimuthal yields for positive and negative helicities $N^+$ and 
$N^-$,

\begin{equation}
	\frac{N^- - N^+}{N^- + N^+} = \frac{A_{e}C_x^{\ast}P_{\gamma}^{\odot} 
	\mbox{sin}\phi_{sc}}{1 + A_{e}P\mbox{cos}\phi_{sc}} .
\label{eq:asym}
\end{equation}

\begin{figure*}[th]
\centerline{
\includegraphics[scale = 0.85]{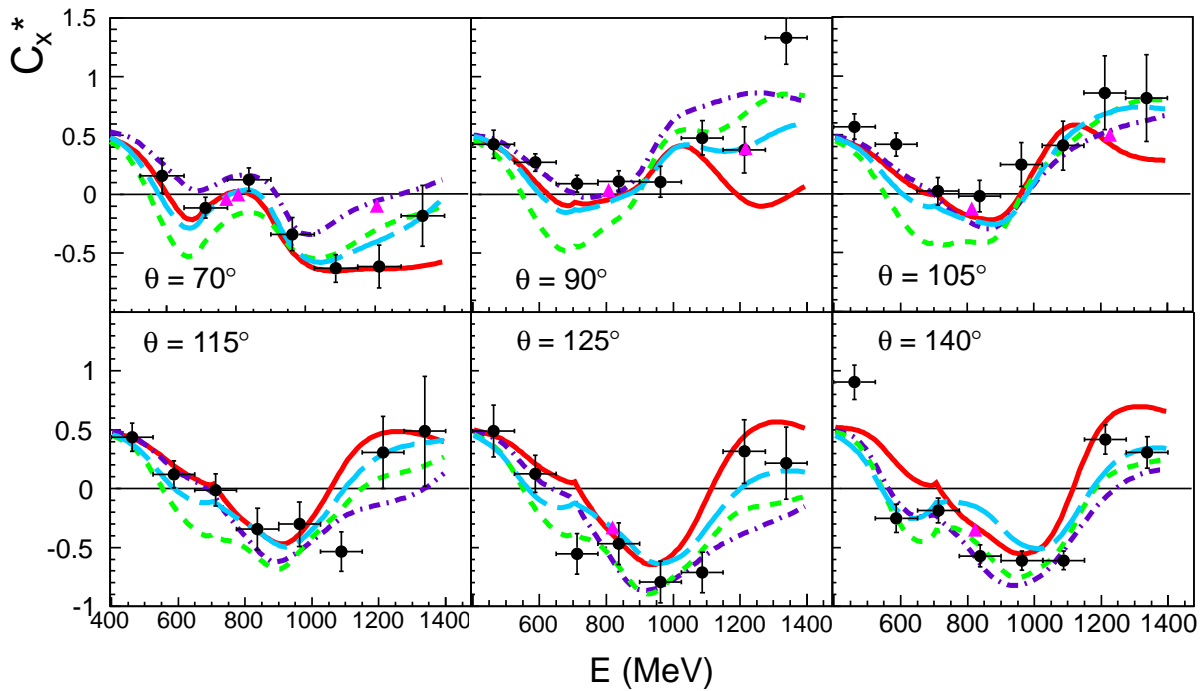}}
\caption{(Color online) $C_x$$^\ast$ excitation functions for
        $\vec{\gamma}p\to\pi^0\vec{p}$ (black circles) for fixed pion CM polar 
        angles. Previous data came from JLab Hall~A~\protect\cite{Gilman} 
        (magenta triangles). The PWA solutions shown are: 
        SAID CM12~\protect\cite{cm12} (cyan long-dash line),
        SAID SN11~\protect\cite{sn11} (green short-dash line), 
        BnGa2011-2~\cite{BnGa} (solid red line), and 
        MAID07~\protect\cite{md07} (violet dash-dotted line). \label{fig:results}}


\end{figure*}

The above expression introduces $A_{e}$, the effective analyzing power 
of the events selected for analysis in that kinematic bin. $A_{e}$ was 
averaged via the simulation over the accepted events. This required a 
realistic parameterization of polarized proton-Carbon scattering. Two 
methods were used for this to provide an estimate of the systematic 
uncertainty. In the first, the world dataset of proton-Carbon 
data~\cite{apow1,apow2,apow3,apow4} was fitted with the parameterization 
given in~\cite{apow4}. In the second, quasi-free scattering, which is 
dominant at these energies, was selected in the simulation and modelled 
using the SAID nucleon-nucleon scattering amplitudes~\cite{NN}, while 
non-quasi-free events were given the former parameterization.
The resulting analyzing powers differed by $\sim$13$\%$. 
This was however found to be the dominant uncertainty 
in the final results. ${A_{e}}$ was obtained from the simulation by 
generating events with $P=0$ and $C_x^{\ast}P_{\gamma}^{\odot}=\pm1$. 
With these conditions the asymmetry, which is fitted to give ${A_{e}}$, 
is:
\begin{equation}
	\frac{N^- - N^+}{N^- + N^+} = A_{e}{\mbox{sin}\phi_{sc}} .
\label{eq:Ae}
\end{equation}

Figure~\ref{fig:comp} demonstrates the excellent agreement between 
the simulated proton scatter angle and the distribution reconstructed 
in the experiment. Below $13^{\circ}$ almost no events undergo a 
nuclear scatter and hence provide little information on the polarization, 
and the width of the distribution reflects the resolution of the angular 
reconstruction. Therefore the polarized scattering simulation was used 
to define the scatter angle acceptance where the figure of merit was 
greatest, i.e., where the statistical uncertainty on ${A_{e}}$, and thus 
the polarization measurement, was minimized. This was done for each 
kinematic bin, yielding the effective analyzing powers shown in Figure~\ref{fig:comp}.

Using the optimized scatter angle cuts described above, the azimuthal 
distribution of accepted scatter events was produced for both real and 
simulated data. The contribution of the analyzing power was then 
removed by dividing the real by simulated distribution. The resulting 
asymmetry is given by dividing eq.~(\ref{eq:asym}) by eq.~(\ref{eq:Ae}) :
\begin{equation}
	A(\phi_{sc}) = \frac{C_x^{\ast}P_{\gamma}^{\odot}} 
	{1 + A_{e}P\mbox{cos}\phi_{sc}} .
\label{eq:fit}
\end{equation}

The circular polarization of the MAMI photon beam $P_{beam}$ can be 
calculated analytically~\cite{PhysRev.114.887} and varied from 30--85\% 
over the $E_{\gamma}$ range of the experiment. Values of $P$ were 
taken bin-by-bin from the current SAID PWA and multiplied by the fits 
to ${A_{e}}$ from the simulation. The extracted $C_x$$^\ast$ results 
show little sensitivity to the value taken for $P$, which gives a 
typical systematic uncertainty of $\sim$0.01. Background processes passing the 
analysis cuts were assessed through simulation to contribute less 
than 3.5\% to the yield.

\section{Results}
\label{sec:results}

In Figure~\ref{fig:results} the new $C_x$$^\ast$ data are 
presented as a function of incident photon energy for a range of 
fixed pion angles alongside the existing measurements from the 
Hall~A Collaboration~\cite{Gilman} at JLab. The data cover a 
center-of-mass energy (W) range of 1277$-$1872~MeV, and therefore 
give constraints on the properties of a large section of the 
nucleon resonance spectrum. The JLab data are limited in kinematic 
coverage because of the small acceptance of the polarimeter but 
are precise. The new data and Hall~A measurements are consistent 
where they overlap, providing a convincing verification of the 
new technique. 

Further interpretation of the results is obtained from comparison 
to the most recent predictions from the SAID PWA. These include the full 
available database of photomeson production reactions and 
meson-nucleon scattering data. Currently two parameterizations 
are available and have been used to extract information on the 
resonance spectrum. From Fig.~\ref{fig:results} it is clear that 
our data are much better described by the newer SAID Chew-Mandelstam 
parameterization, shown as solution CM12, using a technology presented
in Ref.~\cite{cm12}, with $\nicefrac{\chi^{2}}{N}=1.7$, while the SN11 
solution, using a formalism presented in Ref.~\cite{sn11}, gives a 
relatively poor $\nicefrac{\chi^{2}}{N}=3.9$. It should be noted that both these parameterizations give a good description of the previous world database with a $\nicefrac{\chi^{2}}{N}$ of $\sim$2. The differences in the extracted partial wave multipoles from adopting the different parameterizations are reported elsewhere~\cite{cm12} and include significant changes to phase of the $E_{0^{+}}^{1/2}$. The CM12 fit more correctly incorporates the effect of opening quasi-two-body channels above the two-pion threshold. The current MAID~\cite{md07} and Bonn-Gatchina~\cite{BnGa} solutions are also shown on Fig.~\ref{fig:results}. These solutions agree with the overall trends in the new $C_x$$^\ast$ data, although with clear discrepancies for certain kinematic regions. All PWA models will clearly benefit from the constraints provided by this new data.

These results highlight the 
importance of new polarization observables in providing a stringent 
test of PWA methods and in producing new sensitivities, even in kinematic regions where a large number of cross section and polarization observables are already present in the world database. An accurate PWA 
must ultimately describe a complete set of observables. The current data and future experiments exploiting these polarimetry developments at large acceptance detectors will be a key part to achieving this complete measurement. 

\section{Summary and Conclusion}
\label{sec:conc}

A new polarimeter concept has been developed to determine the spin of 
the proton produced in nuclear and hadronic reactions with large 
acceptance and kinematic coverage. This novel, cost effective method 
for large acceptance spin-polarimetry could also find application at 
many other facilities with large acceptance particle detectors such 
as ELSA, JLab, and FAIR where new possibilities for spin observable 
measurements in meson spectroscopy, baryon spectroscopy and nuclear 
structure physics are possible.

In the commissioning experiment a comprehensive set of data for the 
transfer of polarization to the recoiling nucleon ($C_x$$^\ast$) was 
obtained for neutral-pion photoproduction on the proton at incident 
photon energies from 0.4 to 1.4 GeV using the MAMI-C tagged-photon 
beam. The data are a central piece of the ongoing nucleon resonance 
program and give strong evidence that the Chew-Mandelstam formalism 
should be used if reliable information on the excitation spectrum is 
to be obtained.


\vspace{0.5in}
\begin{acknowledgments}
The authors wish to acknowledge the excellent support of
the accelerator group at MAMI. This work was supported
by the Deutsche Forschungsgemeinschaft (SFB 443), the UK Science and
Technology Facilities Council, INFN – Italy, 
the European Community-Research Infrastructure
Activity under FP7 programme (Hadron Physics2,
grant agreement No. 227431),
the Natural Science and Engineering Research Council
(NSERC) in Canada, the National Science Foundation
and Department of Energy in the United States.

\end{acknowledgments}


\end{document}